# The basis of discontinuous motion

Rui Qi

Institute of Electronics, Chinese Academy of Sciences

17 Zhongguancun Rd., Beijing, China

E-mail: rg@mail.ie.ac.cn

We show that the instant motion of particle should be essentially discontinuous and random. This gives the logical basis of discontinuous motion. Since what quantum mechanics describes is the discontinuous motion of particles, this may also answer the question 'why the quantum?'

## 1．Introduction

When most people talk about motion, they only refer to continuous motion, and its uniqueness is taken for granted absolutely but unconsciously. But to our surprise, as to whether continuous motion is the only possible and real motion, no one has given a definite answer up to now.

In classical mechanics, continuous motion is undoubtedly the leading actor. But in quantum mechanics, continuous motion is rejected by the orthodox interpretation from stem to stern. Then why did people never guess that what quantum mechanics describes is another different motion from continuous motion? As we think, this answer is more direct and natural, since classical mechanics describes continuous motion, then correspondingly quantum mechanics, which is different from classical mechanics, may describe another kind of motion. Recently ( Gao Shan [2000a]; Gao Shan [2001] ), a theory of discontinuous motion of particles is presented. It is shown that such motion can be naturally described by the wave function in quantum mechanics, and the simplest nonrelativistic evolution equation of such motion is just the Schrödinger equation. This strongly implies what quantum mechanics describes is discontinuous motion of particles. But why is the motion of particle essentially discontinuous? This need to be further explained. If we can find the logical basis of discontinuous motion, we may also answer the question 'why the quantum?'

In this paper, we will address the above problem. After giving a deep logical analysis about motion, we show that the motion of object should be essentially discontinuous and random. This gives the logical basis of discontinuous motion. We further denote that the evolution law of the discontinuous motion can appear, and when considering gravity it will finally bring about the appearance of continuous motion and Newton's laws in the macroscopic world.

## 2．The objects can move spontaneously

As we know, the object can move or change its position when there is not any outer cause such as outer force. This is an experiential fact, for example, when you kick a ball, it can move freely afterwards. This fact is well summarized in Newton's first law. Besides, there are also other similar phenomena in the microscopic world, for example, the emission of alpha particles by radioactive isotopes happens without outer cause, or we can say, the alpha particles can spontaneously move out from the radioactive isotopes.

On the other hand, there may exist some deep reasons for this counterintuitive fact. One reason is that if the object can't spontaneously move, then the whole world will hold still. As we have known from modern physics, the interaction or force between particles is transferred by the other particles. Now if the particle can't move in a spontaneous way, or it can only move when there is an outer force, then on the one hand, the particle can't move without outer force, on the other hand, the outer force can't exist without the moving particles, which transfer the force. Thus all particles will be motionless, and all forces will not exist. In one word, the whole world will be in a deadly still state.

The direct inference of this conclusion may be that the world will not exist either. Since there is no motion and interaction, the properties of particle, which closely relates to motion and interaction, will disappear, and the particle devoid of any properties will not exist either. Then the world also disappears, and nothing exists[I]. Thus it seems that the objects must move spontaneously in order to exist.

We can define the spontaneous motion of object as the nature or activity of object. Then such nature may be taken as the inner cause for the spontaneous motion of object. Since the nature of spontaneous motion of object doesn't change all the while, this kind of inner cause is independent of time and concrete motion processes.

## 3．How do the objects move spontaneously？

The object can move spontaneously, then how does it move spontaneously? This is a very interesting and important problem. In the following, we will find the instant moving way of the object.

Since the activity of object or the only cause resulting in the spontaneous motion of object is irrelevant to time and concrete motion processes, there is no cause to determine how the object moves spontaneously in space and time. This means that there is no cause to determine the concrete instant motion of the object, i.e. the object is neither determined to move in one special way, nor determined to move in the other special way. Thus the object can only move in a completely random way, or we can say, the instant motion of object must be essentially discontinuous everywhere. As we can see, the reason why the object moves in a random way is just because there isn't any cause to determine a special regular moving way. In short, the object must move, but it doesn't know how to move, so it can only move in a random and discontinuous way.

The above conclusion is also justifiable from a mathematical point of views ( Gao Shan [2000a]; Gao Shan [2001] ). As we know, the motion state of an object in continuous space-time is the infinitesimal interval state, not the instantaneous state. Then the motion state of the object is a point set in space-time[II], but which type of point set is it? According to the mathematical analysis of point set[III], the natural assumption in logic is that it is a general dense point set in space-time,

---

[I] An attractive idea is that the situation in which nothing exists may be logically inconsistent, and can't exist. Then we can interpret the above counterintuitive fact in a complete logical way.

[II] Here the point in the point set represents the mass center of the object at one instant.

[III] As we know, the point set theory has been deeply studied since the beginning of the 20th century. Nowadays we can grasp it more easily. According to this theory, we know that the general point set is dense point set, whose basic property is the measure of the point set. While the continuous point set is one kind of special dense point set, and its basic property is the length of the point set. As an example, as to the point set in two-dimensional

since we have no *a priori* reason to assume a special form, say a continuous point set. Thus during the infinitesimal interval near any instant the object will always move in a random and discontinuous way.

One big obstacle to understand the above conclusion is that people usually think that there exist some laws, say Newton's first law, to determine the existence of a special moving way. Here we will further argue that there doesn't exist such laws at all. Firstly, all laws referring to the time interval, including the infinitesimal time interval, can't determine such moving way. The reason is very simple, since these laws refers to the time-interval motion state of objects, and they are all based on the supposed moving way of objects during the time interval, for example, Newton's first law presupposes the existence of continuous moving way. But in the above discussions, what we consider is the instant motion, not the time-interval state and its evolution, and what we need to find is just the moving way within the time interval. Secondly, physical laws only consider the time-interval motion state of objects. This can be easily seen from the mathematical quantities *dt* and *dx* which appear everywhere in physics. Furthermore, present physics doesn't analyze the way of instant motion, it only supposes the way of instant motion, for example, classical physics presupposes that the instant motion is continuous.

Now as an example, let's see why the instant motion is not continuous. Since what we analyze is the instant motion, the velocity, which is defined on the time interval, doesn't exist yet, and Newton's first law can't help either. Then the free object has no velocity to hold, and it really doesn't know which direction to move along. Thus the object can't move in a continuous way, since continuous motion requires a definite direction, for example, in one-dimensional situation, the object must select a preferred direction, right or left to move continuously.

## 4．The confirmation of discontinuous motion

It seems very strange that the objects move in a discontinuous and random way. But where to find such weird displays? In fact, they exist everywhere in the microscopic world. As one example let's have a look at the well-known double-slit experiment.

In the experiment, the single particle such as photon is emitted from the source one after the other, and then passes through the two slits to arrive at the screen. In this way, when a large number of particles reach the screen, they form the double-slit interference pattern.

Now we will demonstrate that this experiment clearly reveals that the motion of particle is discontinuous. If the motion of particle is continuous, then the particle can only pass through one of the two slits, and it is not influenced by the other slit in each experiment. Thus it is evident that the double-slit interference pattern will be the same as the direct mixture of two one-slit patterns, each of which is formed by opening each of the two slits, since the passing process of each particle in double-slit experiment is exactly the same as that in one of the two one-slit experiments. But all known experiments show that there exist obvious differences between the interference patterns of the above two situations. Thus the motion of particle can't be continuous, and must be discontinuous, especially, the particle must pass through both slits during passing through the two slits. This is an inevitable logical conclusion inferred from the observed double-slit interference pattern if we assume the only existence of particles[I]. But can we directly confirm this conclusion

---

space-time, the general situation is the dense point set, while the continuous curve is one kind of extremely special dense point set. Surely it is a wonder that so many points bind together to form one continuous curve by order-in fact, the probability for its natural formation is zero.

[I] Here Bohm's hidden variables theory doesn't influence the conclusion, in which there are two kinds of existence,

in experiments? The answer is definitely yes ( Gao Shan [2000a], Gao Shan [2000b] ).

As we know, the usual position measurement will destroy the double-slit interference pattern, and can't measure the real state of the particle passing through the two slits. Thus in order to find and confirm the objective discontinuous motion picture of the particle passing through the two slits, we need a new kind of measurement. Fortunately, such measurement method has been found several years ago (Aharonov, Y. and Vaidman, L. [1993]; Aharonov, Y. Anandan, J and Vaidman, L. [1993] ), and its name is protective measurement. According to the principles of protective measurement, since we know the state of the particle beforehand in double-slit experiment, we can protectively measure the objective motion state of the particle when it passes through the two slits. At the same time, the state of the particle will not be destroyed after such protective measurement, and the interference pattern will not be destroyed either. The results of such protective measurement will confirm that the particle passes through both slits, which indicates that the motion of particle is indeed discontinuous.

## 5．The appearance of laws and continuous motion

Even through the instant motion of particle is essentially discontinuous and random, the laws can appear for the interval motion state ( Gao Shan [2000a]; Gao Shan [2001] ). Here we simply introduce the main results. As we have known, the motion state of a particle in continuous space-time is the infinitesimal interval state. Then as to the discontinuous motion, the motion state of particle is a dense point set in space-time, which mathematical description is the position measure density $\rho(x,t)$ and position measure flux density $j(x,t)$. It can be further demonstrated that its simplest nonrelativistic evolution equation is just the Schrödinger equation in quantum mechanics, in which the wave function is the very mathematical complex composed of $\rho(x,t)$ and $j(x,t)$. Furthermore, it can be also shown that, when considering gravity, space-time will be essentially discrete, and the discontinuous motion in discrete space-time may naturally result in the stochastic collapse process of the wave function. This collapse will finally bring about the appearance of continuous motion and Newton's laws in the macroscopic world.

## 6．Conclusions

We solve the problem how the free objects move spontaneously. It is shown that the instant motion of objects should be essentially discontinuous and random, while the continuous motion in the macroscopic world is only apparent. This gives the logical basis of. discontinuous motion, and may answer the question 'why the quantum?'

---

one is the particle, the other is the wave function.